# Spatiotemporal Estimation of TROPOMI $NO_2$ Column with Depthwise Partial Convolutional Neural Network

Authors: Yannic Lops[1], Masoud Ghahremanloo[1], Arman Pouyaei[1], Yunsoo Choi*[1], Jia Jung[1], Seyedali Mousavinezhad[1], Ahmed Khan Salman[1], Davyda Hammond[2]

[1]Department of Earth and Atmospheric Sciences, University of Houston, TX 77004
[2]Oak Ridge Associated Universities, Oak Ridge, TN 37830
*Corresponding author: Yunsoo Choi (ychoi23@central.uh.edu)

**Abstract**
Satellite-derived measurements are negatively impacted by cloud cover and surface reflectivity. These biases must be discarded and significantly increase the amount of missing data within remote sensing images. This paper expands the application of a partial convolutional neural network (PCNN) to incorporate depthwise convolution layers, conferring temporal dimensionality to the imputation process. The addition of a temporal dimension to the imputation process adds a state of successive existence within the dataset which spatial imputation cannot capture. The depthwise convolution process enables the PCNN to independently convolve the data for each channel. The deep learning system is trained with the Community Multiscale Air Quality model-simulated tropospheric column density of Nitrogen Dioxide ($TCDNO_2$) to impute TROPOspheric Monitoring Instrument $TCDNO_2$. The depthwise PCNN model achieves an index of agreement of 0.82 and outperforms the default PCNN models, with and without temporal dimensionality of data, and conventional data imputation methods such as inverse distance weighting by 3-11% and 8-15% in the index of agreement and correlation, respectively. The model demonstrates more consistency in the reconstruction of TROPOspheric Monitoring Instrument tropospheric column density of $NO_2$ images. The model has also demonstrated the accurate imputation of remote sensing images with over 95% of the data missing. PCNN enables the accurate imputation of remote sensing data with large regions of missing data and will benefit future researchers conducting data assimilation for numerical models, emission studies, and human health impact analyses from air pollution.

**Highlights:**

- We use depthwise convolutions within the partial convolutional neural network to confer temporal dimensionality for the imputations of missing remote sensing data.
- We use air quality modeling data and extensive image augmentation to optimize the depthwise partial convolutional neural network.
- The model can accurately impute remote sensing data with large regions of missing data.
- The model outperforms the conventional partial convolutional neural networks for spatiotemporal imputation in both accuracy and consistency.

# Introduction

Nitrogen oxides ($NO_X$=NO+$NO_2$) are some of the major pollutants [1] resulting from human activity [2]. $NO_X$ sources include anthropogenic and natural origins, such as the combustion of fossil fuels [3-4], burning of biomass [5], soil microbial activity [6], and lightning [3]. Nitrogen dioxide ($NO_2$) has been associated with adverse negative health conditions such as cardiovascular diseases [7] and respiratory-related ailments [8-9].

Remote sensing measures the characteristics of an area by utilizing reflected and emitted radiation at a distance. Satellite remote sensing instruments have contributed essential data pertaining to the global distribution [10-12], evolution [13], and the transport of atmospheric pollutants [14-17]. Unfortunately, remote sensing has limitations such as low spatial and temporal resolutions [18] and measurement issues caused by the impact of cloud cover contamination, false reflectance, and significant bias within the data [19-20]. Furthermore, the system can also experience sensor errors that corrupt or lead to failed data measurements [21-22]. Thus limiting the comprehensive application of remote sensing data for forecasting and data assimilation techniques for chemical transport models [2, 4, 23-24].

To non-temporally impute missing data within remote sensing images, studies have applied several methods such as geostatistical approaches [25-26], linear regression models [21], inpainting algorithms [27-28], and deep learning algorithms [29-31]. Deep learning algorithms [32] have shown significant promise in addressing the limitations of missing data, for they model high-level abstractions within datasets [33-34]. Among the various deep learning algorithms, convolutional neural networks (CNNs) [35] have been among the most successful and widely used approaches [36-37] for various purposes such as forecasting [38-42], classification [43-45], speech recognition [46-47], and imputation [30-31,48]. Nevertheless, a number of models and methods still have difficulty in imputing remote sensing data that is missing a significant percentage of the data or contains large gaps within datasets [48-50]. Advanced methods of imputation use temporal dimensionality to enhance the accuracy of the imputation process [49, 51-52].

Although convolution models have been used to impute missing remote sensing data with a temporal dimension within the dataset [48], they require data with a low frequency of missing pixels. This paper expands the application of a partial convolutional neural network (PCNN) [53] to imputing missing remote sensing data [31] by adding temporal dimensionality within the model. The PCNN model performs well at imputing images with a significant amount of missing data and spatial distances, and its performance is further enhanced by the addition of the temporal dimension of the model input. The temporal component of the PCNN is applied through the implementation of depthwise convolutions [54], in which the convolution process is independently performed for each channel. The depthwise partial convolutional neural network (DW-PCNN) aims to address the limitations of the regular PCNN model [31] by incorporating the temporal component for imputation, improving the sharpness of the imputed image and enhancing the accuracy over that of the regular PCNN.

# Methods

## Data Preparation

The TROPOspheric Monitoring Instrument (TROPOMI) is a key instrument aboard the Copernicus Sentinel-5 Precursor (S5P) satellite. The instrument obtains data of key atmospheric constituents such as ozone ($O_3$), $NO_2$, formaldehyde ($CH_2O$), and aerosol through ten spectra bands of ultraviolet (UV), visible (VIS), near-infrared (NIR) and shortwave infrared (SWIR) [55]. The system is a near-polar, sun-synchronous orbit that provides daily global coverage at high spatial resolution (7 km × 3.5 km at the nadir) [56-57].

We performed an initial data filtering process to exclude pixels failing the initial quality assurance (QA) value threshold (i.e., QA<0.5) that presents an error flag or solar zenith angle exceeding 70°, cloud cover, and air mass factor below 0.1 [58]. In addition, we filtered each image to exclude isolated pixel clusters of four or fewer pixels within a defined filter grid to ensure the exclusion of outliers.

We implemented the United States Environmental Protection Agency (U.S. EPA) Community Multiscale Air Quality (CMAQ v5.2) model [59]. The domain of the model has a 12 km grid horizontal spacing with 27 vertical layers reaching 100 hPa, which we used to estimate and predict the tropospheric column density of $NO_2$ ($TCDNO_2$) over the contiguous United States (CONUS). The system uses CB6 and AERO6 chemical mechanisms for the gas-phase and aerosol chemical processes. We also used the 2017 U.S. EPA National Emission Inventory (NEI) [60-61] with parameterized lightning-induced emissions, biogenic emissions computed by using the Biogenic Emission Inventory System, and biomass burning emissions used by the Fire Inventory from the National Center for Atmospheric Research model, version 1.5 [62-64]. The CMAQ model received simulated meteorological variables from the Weather Research and Forecasting model version 4.0 from the National Centers for Environmental Prediction. For initial and boundary conditions, the North Americans Regional Reanalysis (NARR) data were utilized. Furthermore, we incorporated indirect soil moisture and the temperature nudging technique [65-66], as well as a four-dimensional data assimilation option for the temperature, the water vapor mixing ratio, and wind components [67] to enhance the model performance in simulating meteorological fields and performed simulations for the months of February-June, 2019 and 2020.

The simulated $NO_2$ column by CMAQ acts as the basis for preparing training data for the PCNN model. This process ensures that the training data contains images without any missing data. The pixel size of the original CMAQ output images was 299 × 459, and we extracted ten images with a resolution of 256 × 256. To ensure the availability of enough training samples for the partial CNN model, we applied several image augmentation processes [68]. The first image augmentation step of the process was to apply a random noise function with Gaussian smoothing to the CMAQ images to replicate the pixel variations observed in the TROPOMI images. The second step was to apply basic image augmentation to the updated CMAQ image through the random selection of rotation, flipping, or combination. The final augmentation phase involved the random selection, rotation, flipping, and application of TROPOMI masks to the updated CMAQ images. The augmentation phase ensured that the partial convolution had enough training data to become robust at imputing missing data in various images and to ensure that the system did not produce extreme variations or outliers within the missing data imputation process.

The domains of TROPOMI and CMAQ within the CONUS are shown in Figure 1.

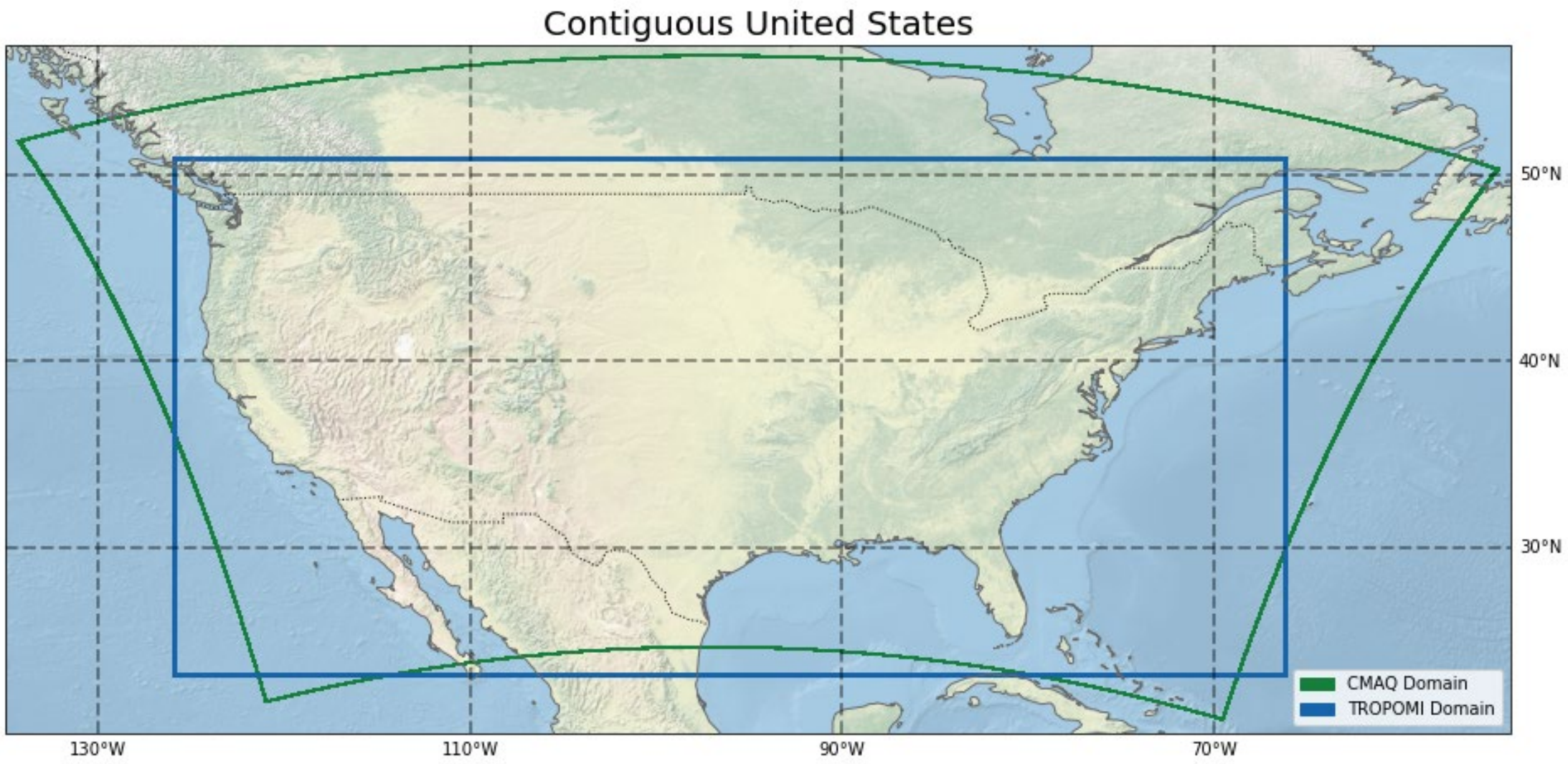

**Figure 1:** Map of the contiguous United States (CONUS) with domains of the CMAQ model (green) and TROPOMI measurement (blue) datasets used for the study.

## DW-PCNN Structure

We used a deep convolutional neural network (deep CNN) [35] based U-net architecture [69] that replaces conventional convolutional layers with partial convolution layers [53]. CNNs process data by convolving the data at each layer over multiple image channels, assigning weights and biases at various aspects within the data, and differentiating between them. The benefits of CNN models are their capability to reduce data into a more manageable form for processing (without losing key features) and extract high-level features of the input data through the use of kernels during the convolving phase [70].

Single-channel images are defined as grayscale images made of one of the primary colors (red, green, and blue) from the three channels. Most common digital images use three channels to collectively form a colored image. As the PCNN model can process one or three channels from an image at a specific instance that the image represents, no temporal dimension is considered. The utilization of recurrent neural networks (RNNs) was considered, but because of the size and complexity of the PCNN model, the utilization of RNNs would have significantly increased the training time of the model without sufficient improvements [38]. Thus, we applied the temporal dimensionality of the PCNN model by including gray-scale images of the TROPOMI images for each channel within the digital image format.

The original partial convolution padding process gradually reduces the significance of the missing data mask at each encoding phase of the PCNN model. Compared to the regular convolution process, the partial convolution process depends only on valid pixels and normalization is adjusted to only a fraction of missing data. During this process, the convolution padding of the mask is applied in unison across all channels. Unfortunately, using different masks for each channel with the conventional convolution kernel causes the convolution kernel to process all the channels at the same time as one unified mask. This process does not reduce the significance of the mask at a gradual pace but at a much faster rate than expected, losing individual features of the mask and reducing the potential performance of the model. To address this limitation, we replaced the conventional 2D convolution layer (within the partial convolution) with depthwise convolutions [54] within the encoding phase of the PCNN model (see Figure 2). A comparison of regular convolution padding and depthwise padding appears in Figure S1. Since the significance of the mask is already removed during the final encoding process of the model, the decoding phase of the U-net architecture remains unchanged. We implemented the PCNN algorithm in the Keras and TensorFlow environments [71-73].

We trained the PCNN model on CMAQ $TCDNO_2$ images of model runs from 2018-2019 and extracted missing data masks from the TROPOMI images. To perform a partial normalization process of the CMAQ TROPOMI images, we divided the entire dataset by a set value of $1\text{x}10^{17}$ based on a value slightly above the maximum $TCDNO_2$ within the TROPOMI 2019 dataset. This ensured that proper distribution or regularization of the data would improve the model performance by reducing the significance of rare outliers within the dataset [40, 74-75]. To further increase the number of training samples and enhance the robustness of the model, we implemented two phases of image augmentation (transforming data into modified samples) [76] of the CMAQ images to generate more training images (see Table S1). The first phase involved applying a modified form of white noise within the CMAQ images to more accurately represent the pixel variations of the original TROPOMI images (see Figure S2). The second phase involved the basic rotation and flipping of the image as well as a randomized linear function that added a positive or negative value shift of the image. TROPOMI $TCDNO_2$ missing data masks also underwent basic augmentation (randomized flipping and rotation) and were overlaid on each CMAQ image.

The partial convolution model structure consists of a total of 16 layers comprised of one input layer, seven depthwise partial convolution encoding layers, seven partial convolution decoding layers, and an output layer. Each encoding layer contains a pooling layer, a depthwise partial convolution layer with batch normalization, and each decoding layer consists of an upsampling layer, concatenated with a respective layer in the encoding layer. We processed the upsampling layer through a partial convolution layer with batch normalization. As the input data contained both negative and positive values, each encoding and decoding layer used the leaky rectified linear unit (ReLU) activation function (negative slope coefficient = 0.5). Leaky ReLU prevents information loss and allows the negative parts of features within the convolution to activate [77-78]. We then processed the final decoding layer through a regular convolutional layer with the hyperbolic tangent (tanh) activation function, which provided the final output image (see Figure 2) for the schematic of the system. We used the same loss function for the model training like that in Liu et al. [53]:

$$L_{total} = L_{valid} + 6L_{hole} + 0.05L_{perceptual} + 120\left(L_{style_{out}} + L_{style_{comp}}\right) + 0.1L_{tv} \quad (1)$$

which is comprised of pixel hole loss ($L_{hole}$), pixel valid loss ($L_{valid}$), perceptual loss ($L_{perceptual}$), raw style output ($L_{style_{out}}$), composited output ($L_{style_{comp}}$), and total variation loss ($L_{tv}$).

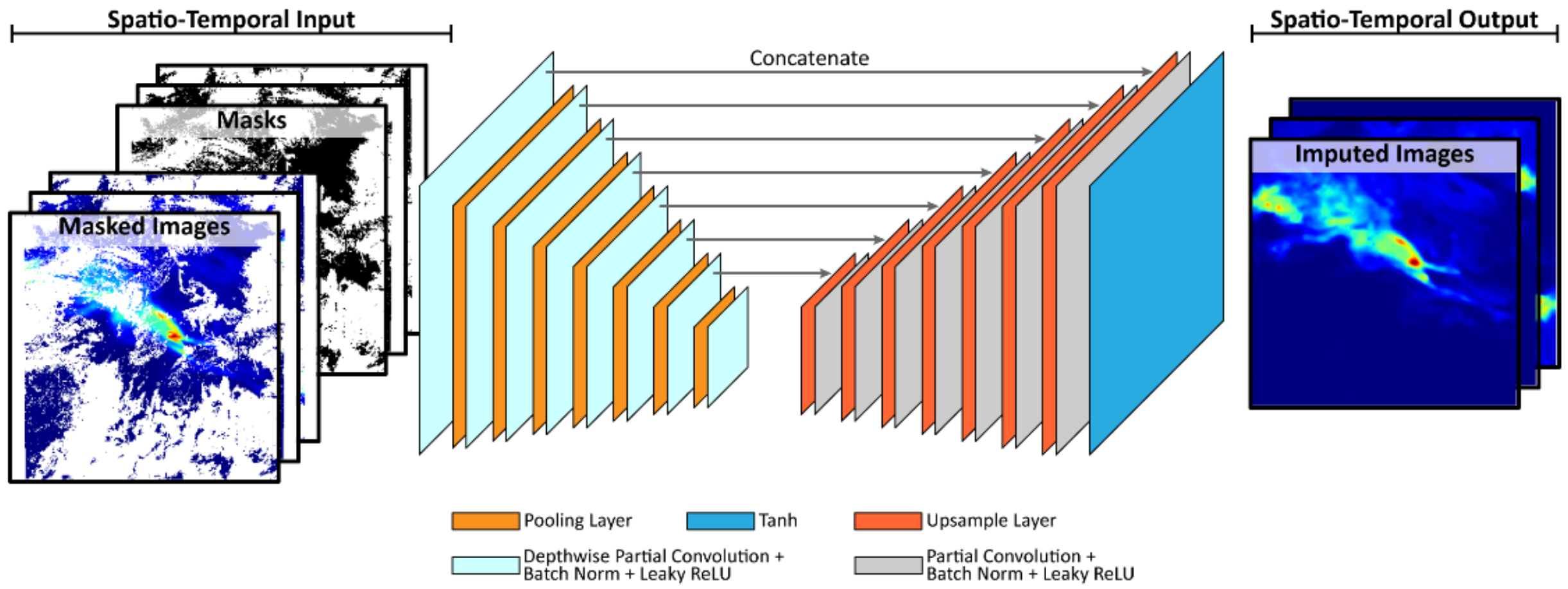

**Figure 2:** Schematic structure of the DW-PCNN model for imputing missing remote sensing data.

The training of the models consisted of three phases: i) the first training phase with batch normalization for 400 epochs with a learning rate of 0.001; ii) the second training phase without batch normalization for 800 epochs with a learning rate of 0.001, and iii) the final training phase with a reduced initialized learning rate of 0.0001 for 800 epochs. Batch normalization is performed during the first training phase to improve the initial speed of the model training by more effectively reducing the loss [79]. The second and third phases exclude batch normalization to optimize loss and reduce the potential bias of the model for imputation. Based on internal tests, the three-phase training improved loss optimization and reduced the overall training time. The optimizer used for the model was the adaptive moment estimation (Adam) [80] stochastic gradient descent method, which adaptively estimates the first- and second-order moments. Checkpoints were enabled by saving the model with the lowest validation error from each training phase.

## Model Comparisons

To compare the performance of the DW-PCNN, we used non-temporal-based imputation methods such as inverse distance weighting (IDW) [81] and the regular PCNN without depthwise convolutions. The IDW interpolation method assumes that pixels close to each other are likely to have similar values and that the local influence of available points on predictions diminishes with distance [82]. These two models showed the best performance in a previous study for the non-temporal imputation of Geostationary Ocean Color Imager aerosol optical depth images [31]. Because of the size of the dataset, which led to significant processing time that required more memory than the high-performance computing system could allocate, implementing the spatiotemporal kriging method [83] as a direct comparison to the DW-PCNN was not possible; therefore, to fill in any remaining missing data, we used IDW for the weekly mean TROPOMI images and integrated the results with CoKriging [83]. The IDW-CoKriging coupled system used the IDW imputed weekly mean (as a substitute to the temporal mean of the dataset) and fed it to the CoKriging process as a co-variable. CoKriging takes advantage of the covariance of the potential relationship of regionalized variables (the weekly mean within filled missing datasets by IDW) during the imputation process. Kriging, based on Gaussian process regression, assumes that spatial variation in a phenomenon is statistically homogeneous throughout a surface based on available data from nearby locations [84]. Both Kriging and IDW (weighting power = 5) models were based on the *gstat* package [85].

## Evaluation

We evaluated the models based on various datasets and methods. Since we trained the PCNN model on CMAQ data, we did not conduct an evaluation based on these data. All daily images had measurements of $TCDNO_2$ values within the 2019 and 2020 study periods. We evaluated the imputation models based on TROPOMI $NO_2$ images by processing the daily TROPOMI images into a weekly moving average of TROPOMI images, and we observed a strong temporal correlation (0.96 *r* for 2019 and 2020) in the weekly averages between daily variations (see Figure S3); thus, we performed a weekly shift format (0.69 and 0.68 *r* for 2019 and 2020, respectively) as input for the spatiotemporal imputation models. We evaluated TROPOMI $TCDNO_2$ by applying TROPOMI $TCDNO_2$ daily missing masks on the weekly mean $TCDNO_2$ images. Then we expected the models to accurately impute the $TCDNO_2$ images. In addition, we applied an estimated pixel distance function based on the distance to the nearest available data point within each image mask. The purpose of this process was to evaluate the imputation bias of the models to the distance of the nearest data variable (see Figure S4).

# Results and Discussion

## Imputation of the TROPOMI Images

The evaluation of TROPOMI NO2 imputation performances utilize the index of agreement (IOA) [86], the correlation coefficient (*r*) [87], the root mean square error (RMSE), and the mean absolute error (MAE) [88] methods. The extraction of evaluation variables is based on the differences between the weekly mean mask and the daily masks for their respective days. The resulting extraction compiles the data into a one-dimensional format to be evaluated by the statistical evaluation methods for each daily dataset. The statistical results are separated between 2019 and 2020 data time periods. Due to IOA incorporating both correlation and bias performance, we used IOA to evaluate the model imputation performance based on the percentage of missing data within the image. The TROPOMI dataset for 2019 was composed of images missing 1 to 20% of data with ~17% of images from 2019 and 30% from 2020 with 10% or more pixels missing from an image (see Figure S5).

Statistical results of the various models and algorithms for the TROPOMI 2019 cases appear in Figure 3 and TROPOMI 2020 cases in Figure 4. For both 2019 and 2020 TROPOMI $TCDNO_2$ images, the DW-PCNN model achieved the best overall performance in the IOA (0.81 for 2019 and 0.82 for 2020) and the *r* (0.74 for both 2019 and 2020). The default PCNN model with spatiotemporal data (PCNN-ST) achieved the lowest MAE ($5.77\times10^{14}$ molecules/$cm^2$) and RMSE ($8.45\times10^{14}$ molecules/$cm^2$) statistical results for 2019, while the DW-PCNN had the lowest MAE ($4.55\times10^{14}$ molecules/$cm^2$) and RMSE ($6.21\times10^{14}$ molecules/$cm^2$) scores for 2020. Overall, the statistical comparisons for 2019 showed minimal differences between IDW and the coupled IDW-CoKriging imputation model (0.15% and 0.04% for the IOA and the *r*, respectively). As a result of the reduced processing resources required, we focused on the IDW model for the 2020 comparisons. Based on the percentage of missing values within the evaluated pixels, the DW-PCNN model outperformed all models. For 2019, it outperformed the other methods in 6% to 34% of images with more than 10% missing data, 10% 14% of images with between 5% and 10% missing data, and 4% to 7% of images with less than 5% missing data. For 2020, the DW-PCNN also outperformed the other methods in 8% to 11% of images with more than 10% missing data, 4% to 11% of images with between 5% and 10% missing data, and 2% to 11% of images with less than 5% missing data. In contrast, the IDW-CoKriging model and the default PCNN models had mixed results with no clear indicator of which model was more accurate overall (see Table S2). Furthermore, both DW-PCNN and PCNN-ST were able to impute a TROPOMI image with over 95% missing data significantly more accurately than the other models (see Figure S6). The advantages and disadvantages of the models based on the results on imputing TROPOMI $NO_2$ are provided in Table S3.

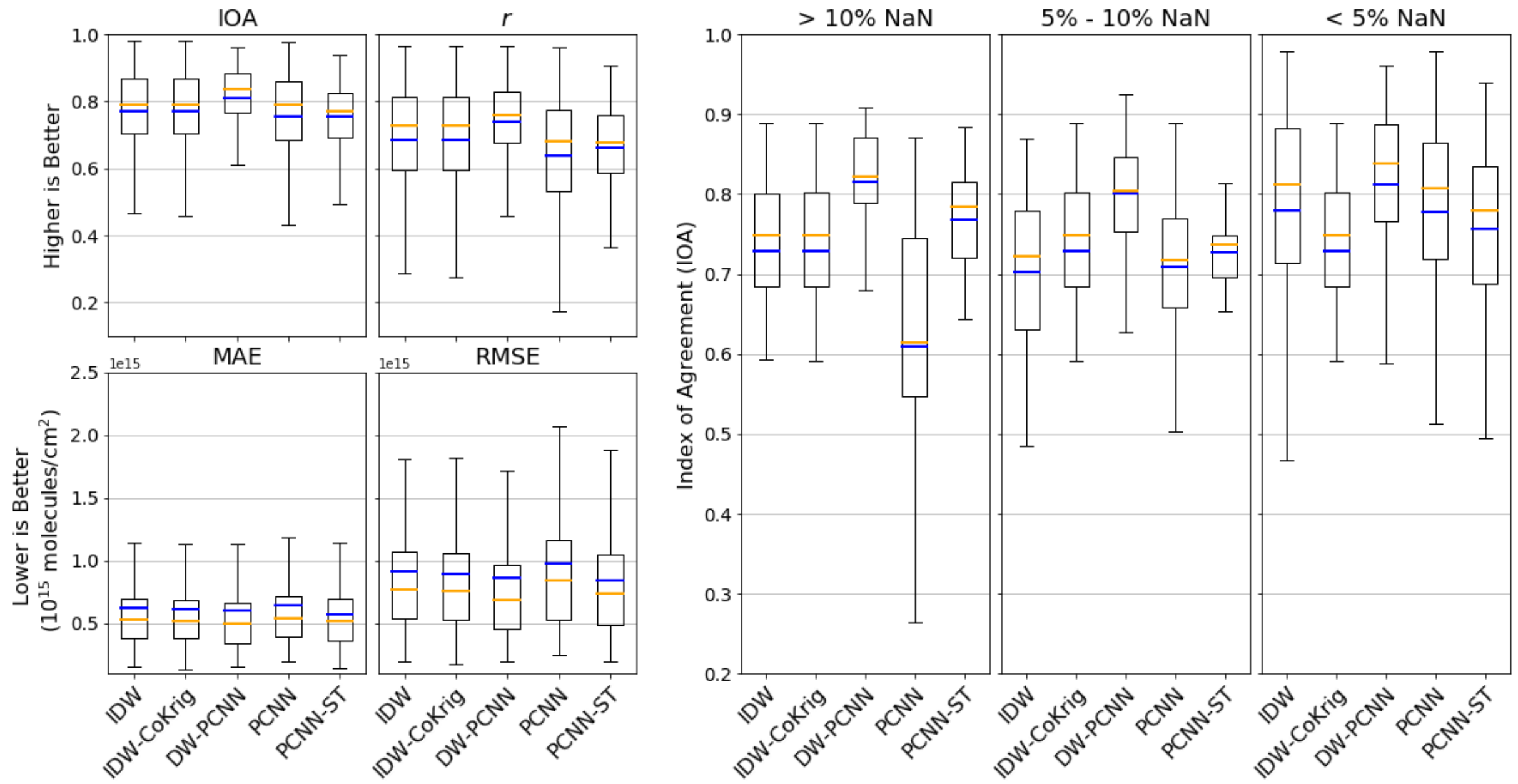

**Figure 3:** Statistical comparison of imputed missing TROPOMI 2019 data from the daily mask to the weekly mean mask. The left figures indicate the statistical performances of inverse distance weighting (IDW), IDW with CoKriging (IDW-CoKrig), the depthwise partial CNN (DW-PCNN), the default partial CNN without spatiotemporal data (PCNN), and the PCNN with spatiotemporal data (PCNN-ST). Evaluations are based on the index of agreement (IOA), the correlation coefficient (*r*), the mean absolute error (MAE), and the root mean squared error (RMSE). The right figure indicates the IOA performance of the models, based on the percentage of missing data split into three categories. The main section of the boxplot presents an interquartile range between the 25th and 75th percentiles. The whiskers (vertical lines) of the boxplot represent the variability outside the interquartile range. The blue and yellow horizontal lines represent the mean and median of the dataset, respectively.

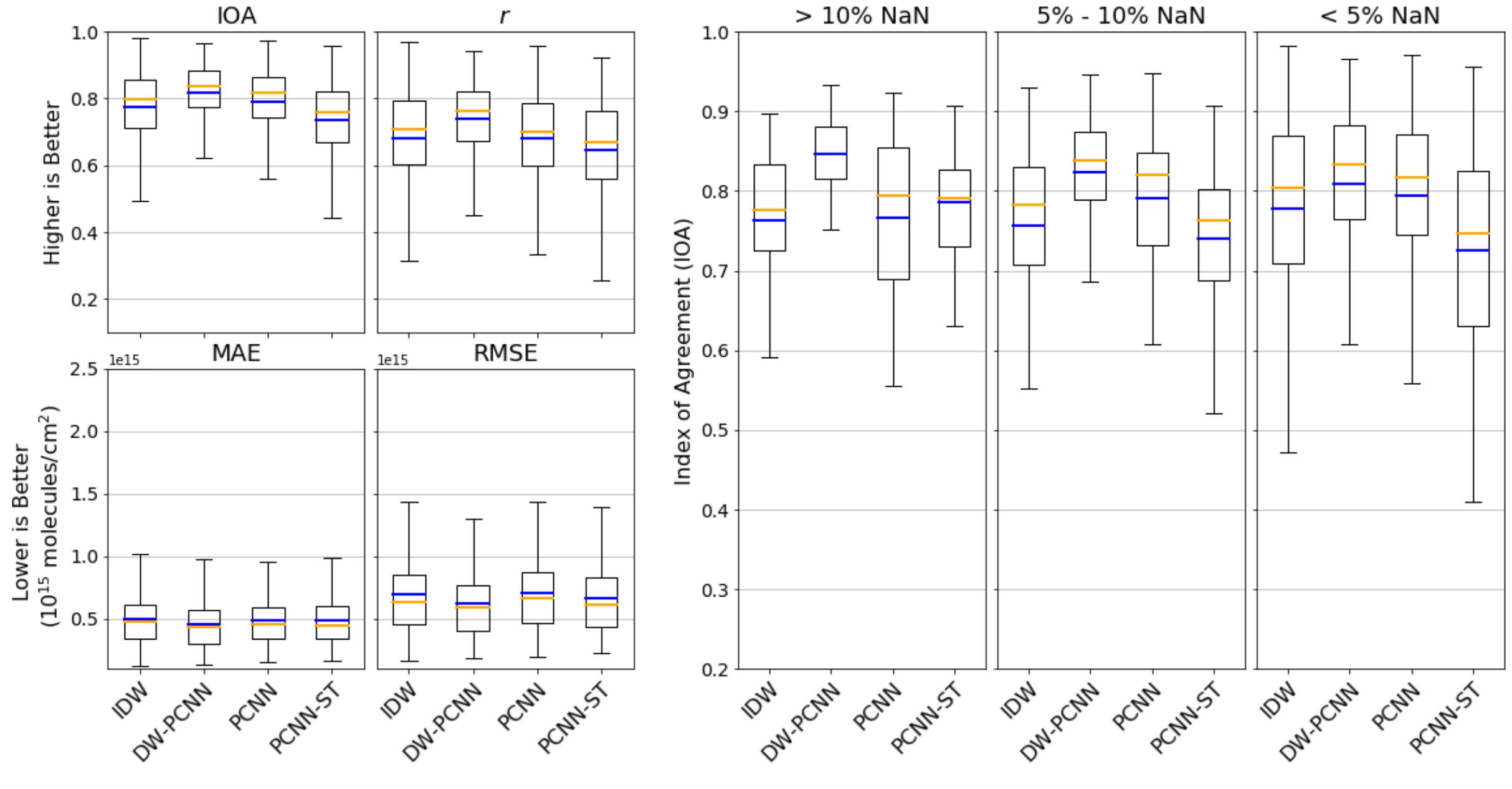

**Figure 4:** Statistical comparison of imputed missing TROPOMI 2020 data from the daily mask to the weekly mean mask. The left figures indicate the statistical performances of inverse distance weighting (IDW), the depthwise partial CNN (DW-PCNN), the default Partial CNN without spatiotemporal data (PCNN), and the PCNN with spatiotemporal data (PCNN-ST). IDW with CoKriging has been excluded due to the minimal overall performance compared to default IDW. Evaluations are based on the index of agreement (IOA), the correlation coefficient (*r*), the mean absolute error (MAE), and the root mean squared error (RMSE). The right figure indicates the IOA performance of the models, based on the percentage of missing data split into three categories. The main section of the boxplot presents an interquartile range between the 25$^{th}$ and 75$^{th}$ percentiles. The whiskers (vertical lines) of the boxplot represent the variability outside the interquartile range. The blue and yellow horizontal lines represent the mean and median of the dataset, respectively.

## Pixel Distance Evaluation

The secondary evaluation method is the comparison of bias in relation to the pixel distance from the nearest available data point within the TROPOMI image. Pixel distance is calculated by applying a Euclidean process for each missing data mask with the output rounded to the nearest integer to represent the pixel distance value. The bias of the imputation methods is performed by the subtraction of the measured pixel to the imputed pixel, which is then categorized based on the distance value of the respective pixels from the pixel distance mask (see Figure S4). These pixel biases within the evaluation mask of each TCDNO2 are then sorted into four-pixel intervals representing ~28 km x ~14 km distances within the TROPOMI data. Each boxplot represents the variances of the biases of the imputed $TCDNO_2$ values to those of the available TROPOMI $TCDNO_2$ value. The series of boxplots at each pixel interval provide information on the range of the bias of each model as the distance increases. We evaluated the DW-PCNN, the IDW CoKriging (for 2019 only), the IDW, the PCNN, and the PCNN-ST models. The pixel distance plot (see Figure 5) shows that the PCNN models had similar variance and ranges of biases throughout the pixel distance range. The PCNN model had a slight negative bias, while the DW-PCNN and the PCNN-ST models had a slightly positive bias. We used IDW CoKriging in the comparison due to the slightly lower overall bias over the default IDW model. IDW CoKriging had an overall negative bias with some fluctuations, but the variance and range of the bias narrowed down more significantly than the PCNN models as the distance increased.

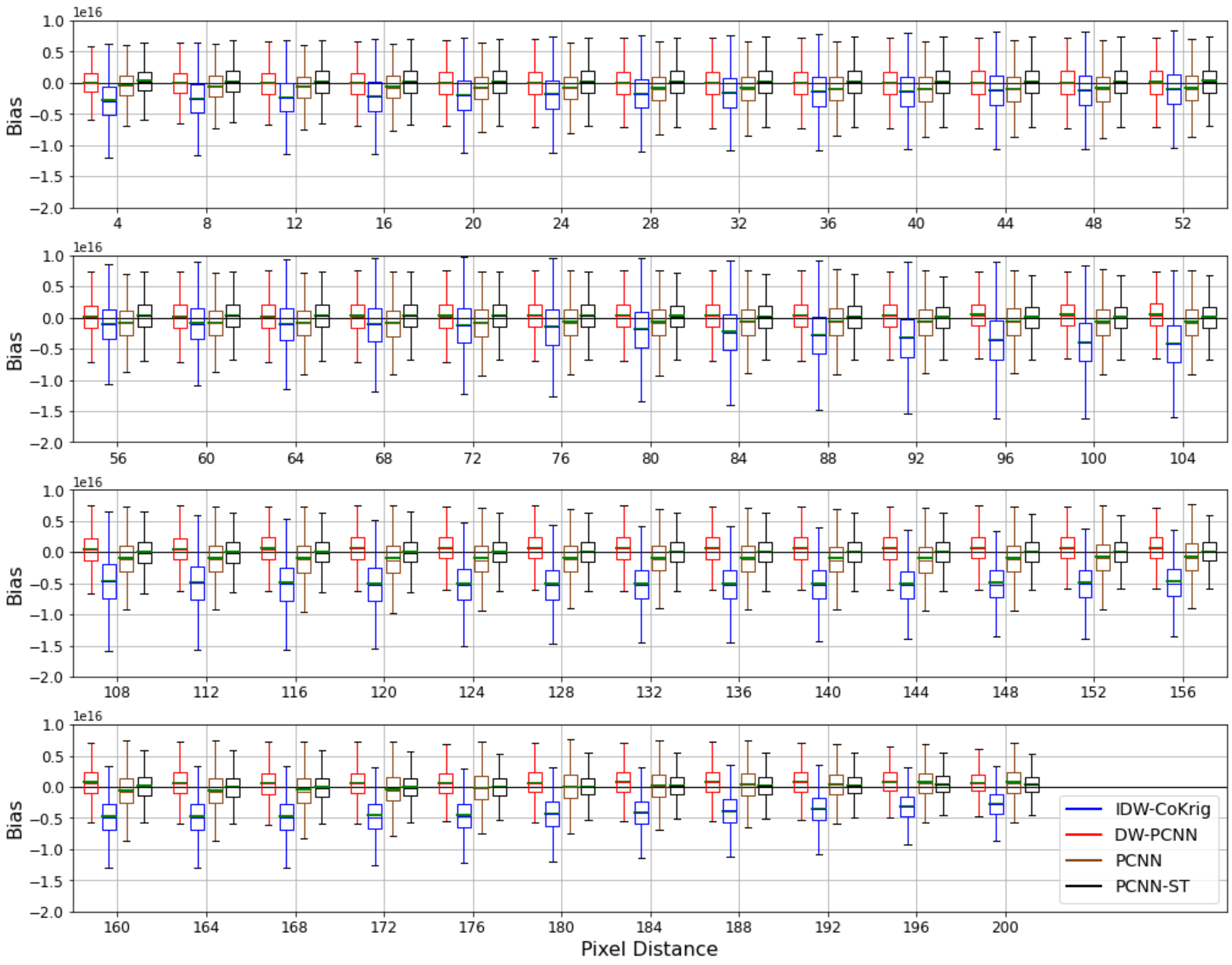

**Figure 5:** Comparison of the bias-variance in the pixel distances of TROPOMI 2019 $NO_2$ of the depthwise Partial CNN (DW-PCNN), coupled inverse distance weighting with CoKriging (IDW CoKrig), default partial CNN without spatiotemporal data as input (PCNN), and default Partial CNN with spatiotemporal data as input (PCNN-ST). The distances are split into four sections at 52-pixel intervals. The main section of the boxplot presents an interquartile range between the $25^{th}$ and $75^{th}$ percentiles. The horizontal lines represent the mean bias of the DW-PCNN (red), the IDW CoKriging (blue), the PCNN (brown), the PCNN-ST (black) models, and the median bias (green for all models) of the imputing missing TROPOMI data. The whiskers (vertical lines) of the boxplot represent the variability outside the interquartile range.

For TROPOMI 2020, the biases of the model imputations were significantly lower than the 2019 biases over distance. The PCNN-ST presented the largest bias within the near pixel distances of the measured TROPOMI data, but compared to the PCNN and DW-PCNN models, it showed a decrease as the pixel distance increased with a slight positive bias. In contrast to the other models, the PCNN experienced an increase in the overall bias and range as the pixel distance increased beyond the 24-pixel (~168 km) distance threshold. IDW also achieved relative stable biases over the distance ranges. Despite the slightly larger bias range compared to the DW-PCNN and the PCNN-ST, the IDW, compared to all of the models, had the most consistent mean bias without a positive or negative trend. The DW-PCNN showed the narrowest bias range of the models across all the pixel distances for the 2020 TROPOMI dataset.

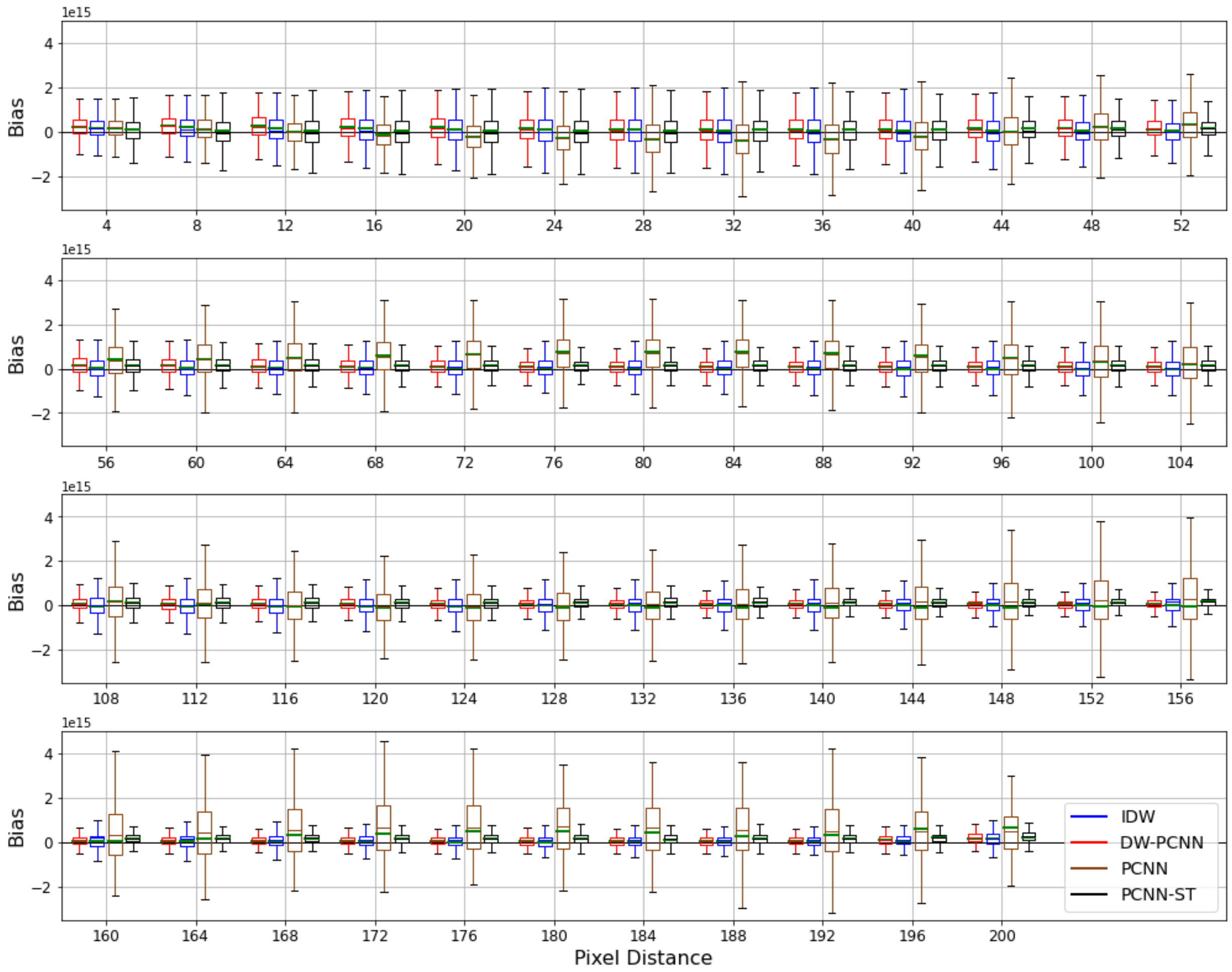

**Figure 6:** Comparison of the bias-variance in the pixel distances of TROPOMI 2020 $NO_2$ of the depthwise partial CNN (DW-PCNN), coupled inverse distance weighting (IDW), default partial CNN without spatiotemporal data as input (PCNN), and default Partial CNN with spatiotemporal data as input (PCNN-ST). The distances are split into four sections at 52-pixel intervals. The main section of the boxplot presents the interquartile range between the 25$^{th}$ and 75$^{th}$ percentiles. The horizontal lines represent the mean bias of the DW-PCNN (red), the IDW (blue), the PCNN (brown), the PCNN-ST (black) models, and the median bias (green for all models) of the imputing missing TROPOMI data. The whiskers (vertical lines) of the boxplot represent the variability outside the interquartile range.

# Conclusion

This research demonstrated the improved capability of the depthwise partial convolutional neural network in the application of spatiotemporal imputation of missing remote sensing data. Both the 2019 and 2020 TROPOMI $TCDNO_2$ imputation results demonstrated that the DW-PCNN most accurately imputed missing TROPOMI $TCDNO_2$ data. With the addition of spatiotemporal data to the PCNN model, it showed significant improvement over the regular PCNN model (without temporal data) with datasets containing large percentages of missing data and at extended distances. Despite the improvements of adding temporal dimensionality within the input of the PCNN model, the mask padding of the regular convolution process shows limitations, which has led to some bias. The implementation of depthwise convolutions, in which the masks at each image channel are padded separately, showed further improvement, demonstrating the importance of maintaining the individual channel masks and gradual feature reduction over the conventional method. Furthermore, the DW-PCNN was the only model that maintained a mean IOA above 0.8 in all of the statistical comparisons for both the TROPOMI 2019 and 2020 $TCDNO_2$ datasets. The current limitation of the DW-PCNN is the constrained number of channels (only three) that the model can process, thus restricting the temporal samples and the addition of co-variables that enhance the accuracy of imputation.

While the bias-distance comparison showed that the PCNN-ST performed as well as the DW-PCNN, the statistical comparisons showed that the PCNN-ST had a lower correlation when reconstructing the TROPOMI $TCDNO_2$ dataset than the default PCNN model. This finding can be explained by the output of the PCNN-ST model, which showed smoother transitions than the PCNN or DW-PCNN models, a slight under-prediction of high $TCDNO_2$ concentrations and an over-prediction of low $TCDNO_2$ column concentrations. Although this phenomenon may minimally impact the evaluation of bias, it impacts the correlation and IOA scores to a greater extent.

For a spatial imputation algorithm without temporal dimensionality, IDW performed consistently compared to the PCNN and PCNN-ST models. Although IDW was not able to surpass the PCNN models in different metrics, it did not perform the worst in the respective metrics. The major limitation of IDW is the computational cost of such large datasets, especially when required to take all available samples within the TROPOMI image. In fact, IDW and other interpolation-based algorithms (e.g., Kriging) require exponentially more computation power and resources as the dataset size increases spatially and when temporal dimensionality is added. For smaller dataset sizes, however, these algorithms show performances similar to that of the PCNN model (refer to [31]). As datasets increase in size, dimensionality, and missing data, the benefits of deep learning algorithms for imputation purposes also increase.

Once trained, not only does the DW-PCNN model impute large remote sensing datasets in significantly less processing time than interpolation-based algorithms, but it is also significantly more accurate than the default PCNN models with and without the temporal dimensionality of datasets. For the accurate imputation of such datasets, the implementation of DW-PCNN, which enables the accurate imputation of remote sensing data with large regions of missing data, will benefit future researchers that conduct studies entailing data assimilation for numerical models, emission studies, and human health impact analyses from air pollution. To further enhance the imputation accuracy of the DW-PCNN model, we need to expand the number of image channels that the PCNN can receive. In addition, if we increase the number of input channels and add more temporal information and co-variables as input, the imputation capability of the model will further improve.

## Acknowledgments

This study was supported by funding from the Oak Ridge Associated Universities (ORAU) Directed Research and Development. The authors acknowledge the Research Computing Data Core (RCDC) at the University of Houston (https://uh.edu/rcdc/) and Texas Advanced Computing Center (TACC) at the University of Texas at Austin (http://www.tacc.utexas.edu) for providing high-performance computing resources that have contributed to the research results reported within this paper. We would like to thank Mathias Gruber for reconstructing the partial CNN code for the TensorFlow implementation, which we have updated and modified. The original code is available at https://github.com/MathiasGruber/PConv-Keras.

**Conflict of interest**

The authors declare that they have no conflict of interest.